%% file: FACT_COLSUP.tex
\input jrmacros.tex

\refstyleprl
\input psfig

\input reforder
\input tables
\centerline{FACTORIZATION AND COLOR-SUPPRESSION}
\centerline{IN HADRONIC $B \rightarrow D^{(*)}n\pi$ DECAYS}
\vskip 24pt

{\singlespace 
\centerline{Jorge L. Rodriguez}
\centerline{{\it Department of Physics and Astronomy, University of Hawaii}}
\centerline{{\it Honolulu, Hi 96822, U.S.A}}
\centerline{{\it University of Hawaii preprint UH 511-842-96}}
}

\def\a2oa1{$a_2/a_1$}

\xdef\chapterno{1}

\subhead{Introduction}
We discuss recent results on hadronic \B decays using data obtained
with the CLEOII detector\refto{CLEOnim}. The results are used to test
the factorization hypothesis and color suppression.

Two body hadronic decays which involve the quark level transition $b\to
c\bar{u}d$ fall into three general categories. Class I and class II
decays involve neutral \B mesons which decay to a charged $D^{(*)}$ 
and a charged meson or to a neutral $D^{(*)}$ and a neutral meson. In these decays the transitions
are mediated by external or internal (color-suppressed) diagrams,
respectively. In class III decays a charged \B decays to a neutral
charmed meson plus a light hadron where the final state quark configuration
can result from either spectator diagrams. In the usual theoretical
treatment\refto{BSW,Neubert,Deandrea,Ruckl} the decay amplitudes are
expressed as linear functions of parameters $a_1$ and $a_2$, each
assigned to the amplitude associated with the external and internal
diagram, respectively.

\subhead{Experimental Procedure}
We have measured branching fractions for five class I decays: $\B^0
\to D^{(*)+}\pi^-,$ $D^{(*)+}\rho^-$ and $D^{*+}a_1^-$ and five class
III modes $\B^- \to D^{(*)0}\pi^-$, $D^{(*)0}\rho^-$ and $\B^- \to
D^{*0}a_1^-$ decays. The data sample used consists of $2.04\ fb^{-1}$
of data collected at the \USSSS by CLEO at the CESR $e^+e^-$ ring. The
sample corresponds to $2\times 10^6$ \bbar pairs. To determine the
event yields we fully reconstructed the decay chains in their
exclusive modes, formed beam-constrained mass distributions and then
fit these to a Gaussian plus a background shape. The combined
beam-constrained mass plots are shown in Figure 1. The various
criteria used in selecting particle candidates used are described in
greater detail in Refs:\refto{BigB,JRthesis}. The branching fraction
measurements obtained are listed in Table 1.

\topinsert
\vskip -1.2truecm
\centerline{\psfig{figure=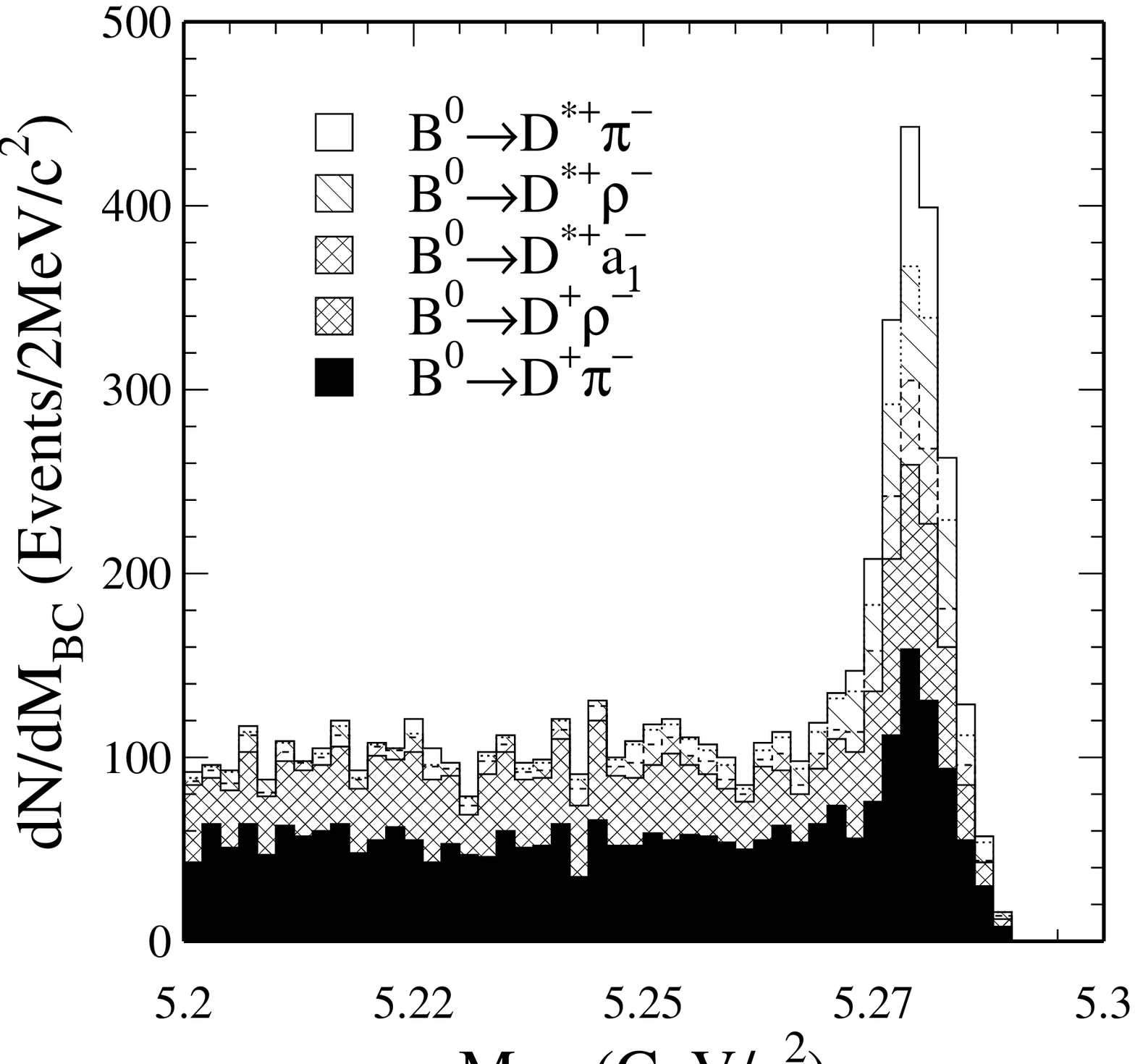,width=7.5cm}\psfig{figure=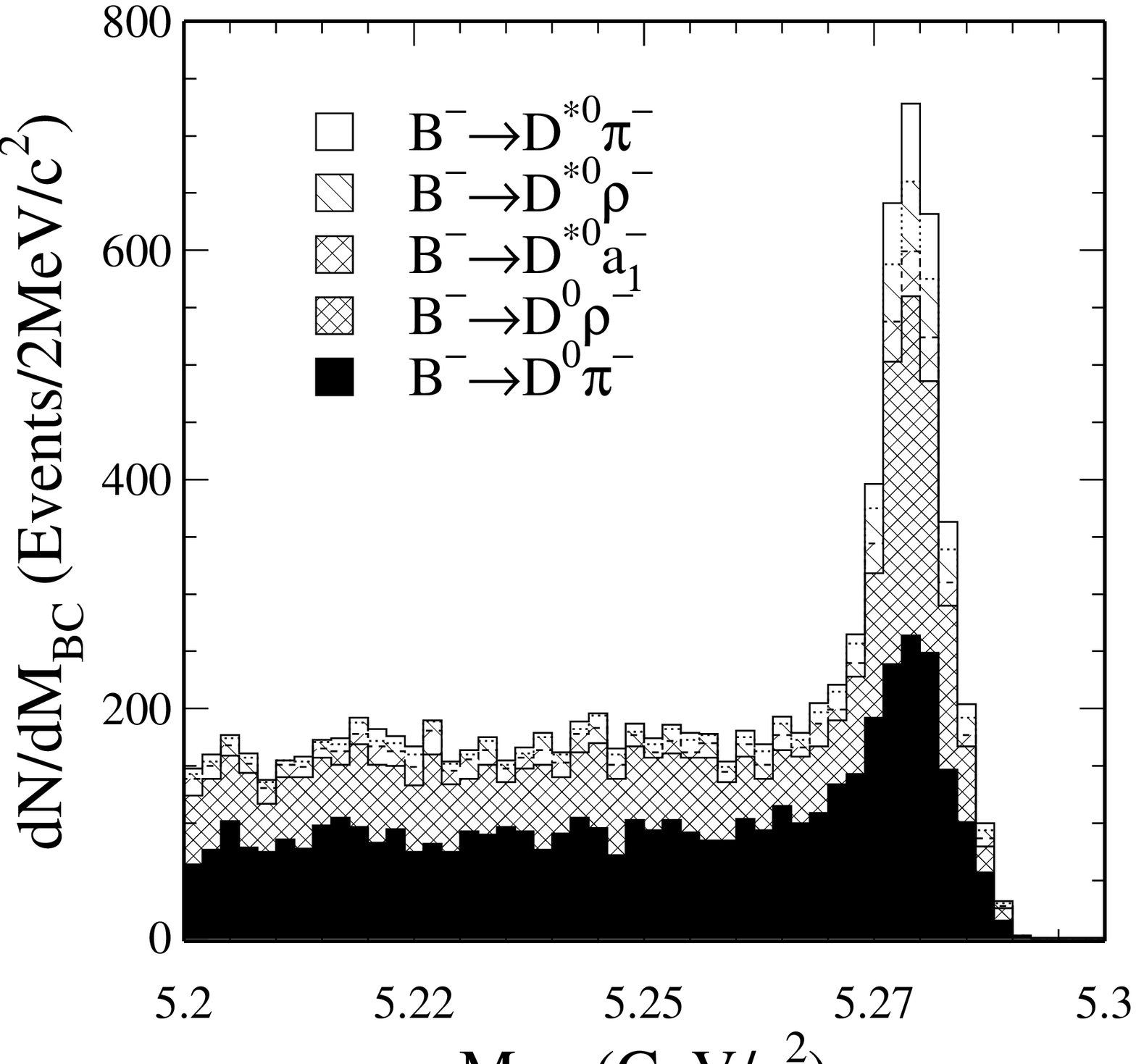,width=7.5cm}}
\vskip 10pt
\figureCaption\BCmassplots{Beam-constrained mass plots.}{}
\endinsert

\subhead{Determination of $|a_1|$ and the relative sign of $a_2/a_1$} 
To determine the values of $a_1$ and $a_2/a_1$ we use the
branching fraction measurements in Table 1 and theoretical
predictions for the branching fractions. The branching fractions of
the first four class I decays are used as inputs in a least squares
fit to obtain the following results
$$
\eqalign{
|a_1| = &1.14 \pm 0.024 \pm 0.022 \pm 0.050 \quad {\rm BSW II} \cr
|a_1| = &1.06 \pm 0.023 \pm 0.021 \pm 0.046 \quad {\rm CDDFGN}. \cr
}
\eqno(a1val)
$$
The first error is statistical, the second is the systematic error and
the third is the error due to the uncertainty in the \B lifetime
and production ratio\refto{Dstrlnu}. The two models used,
BSWII\refto{Neubert} and CDDFGN\refto{Deandrea} employ Heavy 
Quark Effective Theory but differ mainly in the assumption used to
model the $q^2$ dependence of form factors.  

The magnitude of the fit parameter $a_1$ is consistent with
the expectation from QCD and factorization. In class I decays the QCD
coefficients which multiply the matrix elements are given by 
$a_1'=c_1(\mu)+{1\over N_c} c_2(\mu)$. Using NLLA results for
$c_1(\mu)$ and $c_2(\mu)$, at $\mu=5$ GeV and setting $N_c=3$ gives
$a_1'=1.04$.  

To determine $a_2/a_1$ we form ratios of class III to class I decays
and compare the results to theoretical model predictions. The values
in Equation (2) are also determined by performing a least squares fit
to data.
$$
\eqalign{
{a_2\over a_1} =& +0.15 \pm 0.036 \pm 0.047 \pm^{0.107}_{0.084} \quad {\rm BSW II}\cr
{a_2\over a_1} =& +0.16 \pm 0.035 \pm 0.040 \pm^{0.096}_{0.076} \quad {\rm CDDFGN}\cr
}\eqno(a2overa1)
$$
The positive sign of \a2oa1 differs from the expectation obtained by
extrapolating the charm results to the \B system. However, the sign is
consistent with QCD and factorization (with small non-factorizable
contributions). 

\subhead{Direct Tests of Factorization}
To test factorization directly we make use of the fact that in this
approximation hadronic amplitudes are products of two independent
matrix elements. The matrix element describing the heavy to heavy
transition is identical to that in the semileptonic transition while
the production of the light meson from the vacuum can be described by
a simple expression involving numerical and decay constants. To
perform direct tests of factorization we thus check that Equation (3)
$$
\frac{\Gamma\left(\bar{B}^0\to D^{*+} h^-\right)}{\frac{d\Gamma}{dq^2}
\left(\bar{B}^0\to D^{*+} l^- \bar{\nu_{l}}\right)\biggr|_{q^2=m^2_{h}}} =
6\pi^2{ a_1'^2}
f_{h}^2|V_{ud}|^2 
\eqno(Factrela)
$$
is satisfied. The denominator in the LHS is determined, at each $q^2$,
by interpolating the differential $q^2$ spectrum of the semileptonic
decay widths\refto{BigB}. The values for $f_h$ and $V_{ud}$ are taken
from recent experimental results\refto{Weinstein,PDG}. The comparison
between the LHS $(R_{exp})$ and the RHS $(R_{th})$ is given in Table
2.  These show consistency with factorization to present experimental
precision.

A more subtle test of factorization\refto{Kroner} can be performed by
comparing the polarization of final states in hadronic decays to the
polarization in semileptonic decays. We have measured the fraction of
\Dstrrho decays which are polarized in the longitudinal direction to
be $\Gamma_L/\Gamma = 90.0 \pm 3.7 \pm 4.5\%.$ The longitudinally
polarized fraction of semileptonic decays at $q^2=m_\rho^2$ is $85\%$
which is in agreement our with results. The semileptonic value is
extracted by fitting the differential $q^2$ spectrum to model
estimates\refto{Neubertpol}.

\subhead{Color-Suppression}
Class II decays are defined as decays which can proceed only through
internal spectator diagrams. These processes are products of
the effective neutral term which gets multiplied by the scale dependent
$a'_2$. In class II decays the value of $a'_2$ is significantly smaller
than $a'_1$ since it involves the difference of two numbers of similar
size. We thus expect class II decays will be suppressed relative to
class I decays.  

To search for color-suppression we use the large sample of \B
mesons available and examine 10 modes: $B^0\to D^{(*)0}m^0$, where
$m^0=\pi^0,\eta,\eta^\prime,\rho^0$ or $\omega$. The procedure used to
find the event yield was identical to that used for class I and class
III modes. However, since no clear signals were obtained, limits
were set for the branching fractions at the $90\%$ confidence
level. The results are listed in Table 3.

\subhead{Conclusion}
By comparing hadronic decays allowed only through external diagrams to
the corresponding semileptonic decay we show that to current
experimental precision, decays of the type $B^0 \to D^{(*)+}(n\pi)^-$
are consistent with the factorization hypothesis. Further evidence for
factorization is suggested by both the sign of \a2oa1 and magnitude
$a_1$ when compared to expectations from QCD with
factorization. Finally, we show that color-suppression is operative in
$b\to c\bar{u}d$ type transitions with our limits on class II decays.

\subheadnn{Acknowledgments}
I thank my colleagues on the CLEO II experiment for their contribution
to the work presented here. I'd like to also thank the CESR staff for
providing CLEO with record breaking luminosity and enabling CLEO
to accumulate the large statistics reported on in this analysis.

\vskip 24pt
\input tbl_normall.tex

\input tbl_factestbr.tex

\input tbl_color_all.tex

\vskip 24pt

\eject

\references
\nopagenumbers

\refis{Weinstein} A. J. Weinstein, private communication and CLEO
internal note.

\refis{JRthesis} J. L. Rodriguez, University of Florida Dissertation (1995).

\refis{BigB} CLEO Collaboration, M. S. Alam \etal \prd, 50 43 (1994).

\refis{Neubert} M. Neubert, V. Rieckert, B. Stech and Q. P. Xu in 
{\it Heavy Flavours} edited by A. J. Buras and H. Lindner World
Scientific, Singapore (1992). 

\refis{BSW} M. Bauer, B. Stech and M. Wirbel, Z. Phys. {\bf C29} 639 (1985).

\refis{Deandrea} A. Deandrea, N. Di Bartolomeo, R. Gatto and
G. Nardulli, \pl, {B 318} 549 (1993).

\refis{PDG} Particle Data Group, M. Aguilar-Benitez \etal, \prd, 50
1173 (1994).  

\refis{Ruckl} R. R\"uckl, ``Flavor Mixing in Weak Interactions,''
in {\it Ettore Majorana International Science Series}, Ling-Lie Chau
editor, Pleum Press, New York (1984).

\refis{CLEOnim} CLEO collaboration, Y. Kubota \etal, Nucl. Instr. and Meth. 
{\bf A320} 66 (1992).

\refis{Kroner} J. K\"orner and G. Goldstein, \pl, 89B 105 (1979).

\refis{Neubertpol} M. Neubert, \pl, {B 264} 455 (1991).

\refis{Dstrlnu} CLEO Collaboration, B. Barish \etal \prd, 51 1014
(1995).

\endreferences

\bye

%% file: jrmacros.tex


\font\twelverm=cmr10  scaled 1200   \font\twelvei=cmmi10  scaled 1200
\font\twelvesy=cmsy10 scaled 1200   \font\twelveex=cmex10 scaled 1200
\font\twelvebf=cmbx10 scaled 1200   \font\twelvesl=cmsl10 scaled 1200
\font\twelvett=cmtt10 scaled 1200   \font\twelveit=cmti10 scaled 1200
\font\twelvesc=cmcsc10 scaled 1200  


\font\fourteenrm=cmr10 scaled 1440 
\font\fourteenmit=cmmi10 scaled 1440

\skewchar\twelvei='177   \skewchar\twelvesy='60

\font\twelvemib=cmmib10 scaled 1200
\font\tenmib=cmmib10
\font\eightmib=cmmib10 scaled 800

\skewchar\twelvemib='177
\newfam\mibfam
\newfam\xivmitfam


\def\twelvepoint{\normalbaselineskip=12.8pt plus 0.1pt minus 0.1pt
  \abovedisplayskip 12.8pt plus 3pt minus 9pt
  \belowdisplayskip 12.8pt plus 3pt minus 9pt
  \abovedisplayshortskip 0pt plus 3pt
  \belowdisplayshortskip 7.2pt plus 3pt minus 4pt
  \def\rm{\fam0\twelverm}          \def\it{\fam\itfam\twelveit}%
  \def\sl{\fam\slfam\twelvesl}     \def\bf{\fam\bffam\twelvebf}%
  \def\mit{\fam 1}                 \def\cal{\fam 2}%
  				   \def\tt{\twelvett}
  \def\sc{\twelvesc}
  \def\mib{\fam\mibfam\twelvemib}  
  \textfont0=\twelverm   \scriptfont0=\tenrm   \scriptscriptfont0=\sevenrm
  \textfont1=\twelvei    \scriptfont1=\teni    \scriptscriptfont1=\seveni
  \textfont2=\twelvesy   \scriptfont2=\tensy   \scriptscriptfont2=\sevensy
  \textfont3=\twelveex   \scriptfont3=\twelveex  \scriptscriptfont3=\twelveex
  \textfont\xivmitfam=\fourteenmit \scriptfont\xivmitfam=\twelvemib 
                        \scriptscriptfont\xivmitfam=\tenmib 
  \textfont\mibfam=\twelvemib \scriptfont\mibfam=\tenmib
		\scriptscriptfont\mibfam=\eightmib 
  \textfont\itfam=\twelveit
  \textfont\slfam=\twelvesl
  \textfont\bffam=\twelvebf \scriptfont\bffam=\tenbf
  \scriptscriptfont\bffam=\sevenbf
  \normalbaselines\rm}



\mathchardef\alpha="710B
\mathchardef\beta="710C
\mathchardef\gamma="710D
\mathchardef\delta="710E
\mathchardef\epsilon="710F
\mathchardef\zeta="7110
\mathchardef\eta="7111
\mathchardef\theta="7112
\mathchardef\iota="7113
\mathchardef\kappa="7114
\mathchardef\lambda="7115
\mathchardef\mu="7116
\mathchardef\nu="7117
\mathchardef\xi="7118
\mathchardef\pi="7119
\mathchardef\rho="711A
\mathchardef\sigma="711B
\mathchardef\tau="711C
\mathchardef\phi="711E
\mathchardef\chi="711F
\mathchardef\psi="7120
\mathchardef\omega="7121
\mathchardef\varepsilon="7122
\mathchardef\vartheta="7123
\mathchardef\varpi="7124
\mathchardef\varrho="7125
\mathchardef\varsigma="7126
\mathchardef\varphi="7127


\def\beginlinemode{\endmode
  \begingroup\parskip=0pt \obeylines\def\\{\par}\def\endmode{\par\endgroup}}
\def\beginparmode{\endmode
  \begingroup \def\endmode{\par\endgroup}}
\let\endmode=\par
{\obeylines\gdef\
{}}
\def\singlespace{\baselineskip=\normalbaselineskip}

\def\oneandahalfspace{\baselineskip=\normalbaselineskip
  \multiply\baselineskip by 3 \divide\baselineskip by 2}
\def\doublespace{\baselineskip=\normalbaselineskip \multiply\baselineskip by 2}

\newcount\firstpageno
\firstpageno=-100
\footline={\ifnum\pageno<\firstpageno{\hfil}\else{\hfil\twelverm\folio\hfil}\fi}
\def\toppageno{\global\footline={\hfil}\global\headline
  ={\ifnum\pageno<\firstpageno{\hfil}\else{\hfil\twelverm\folio\hfil}\fi}}
\def\nopagenumbers{\global\headline={\hfil}\global\footline={\hfil}}
\def\botpageno{\global\headline={\hfil}\global\footline
  ={\ifnum\pageno<\firstpageno{\hfil}\else{\hfil\twelverm\folio\hfil}\fi}}
\newcount\headpage
\headpage=0
\def\spam{{\hfil\twelverm\folio\hfil}}

\def\updownpages{ 
\global\headline={\ifnum\headpage=\pageno{\hfil}\else\spam\fi}
\global\footline={\ifnum\headpage=\pageno\spam\else{\hfil}\fi}
}

\let\rawfootnote=\footnote		
\def\footnote#1#2{{\rm\singlespace\parindent=17pt\parskip=0pt
  \rawfootnote{#1}{#2\hfill\vrule height 0pt depth 6pt width 0pt}}}
\def\raggedcenter{\leftskip=4em plus 12em \rightskip=\leftskip
  \parindent=0pt \parfillskip=0pt \spaceskip=.3333em \xspaceskip=.5em
  \pretolerance=9999 \tolerance=9999
  \hyphenpenalty=9999 \exhyphenpenalty=9999 }
\def\dateline{\rightline{\ifcase\month\or
  January\or February\or March\or April\or May\or June\or
  July\or August\or September\or October\or November\or December\fi
  \space\number\year}}
\def\received{\vskip 3pt plus 0.2fill
 \centerline{\sl (Received\space\ifcase\month\or
  January\or February\or March\or April\or May\or June\or
  July\or August\or September\or October\or November\or December\fi
  \qquad, \number\year)}}

\hsize=16.5truecm
\hoffset=.0truein 

\vsize=24.9truecm
\voffset=-1.04truecm
\parskip=0pt     
\raggedbottom    
\def\\{\cr}
\twelvepoint		
\oneandahalfspace	
\overfullrule=0pt	
\widowpenalty=10000         
\clubpenalty=10000	    


\gdef\pgtoskip{6}		
\gdef\npgtoc{3}			
\gdef\npgtotbl{2}		
\gdef\npgtofig{4}		

\global\newcount\toallpageno
\global\newcount\tocskip		
\global\newcount\totblskip		
\global\newcount\tofigskip		

\global\tocskip=\pgtoskip
\global\totblskip=\pgtoskip \global\advance\totblskip by \npgtoc
\global\tofigskip=\totblskip \global\advance\tofigskip by \npgtotbl
\global\toallpageno=\tofigskip \global\advance\toallpageno by \npgtofig
%



\newcount\timehour
\newcount\timeminute
\newcount\timehourminute
\def\daytime{\timehour=\time\divide\timehour by 60
  \timehourminute=\timehour\multiply\timehourminute by-60
  \timeminute=\time\advance\timeminute by \timehourminute
  \number\timehour:\ifnum\timeminute<10{0}\fi\number\timeminute}
\def\today{\number\day\space\ifcase\month\or Jan\or Feb\or Mar
  \or Apr\or May\or Jun\or Jul\or Aug\or Sep\or Oct\or
  Nov\or Dec\fi\space\number\year}



\def\title			
  {\null\vskip 3pt plus 0.2fill
   \beginlinemode\twelverm\singlespace \raggedcenter }

\def\author			
  {\vskip 3pt plus 0.2fill \beginlinemode
   \twelverm\doublespace \raggedcenter}

\def\affil			
  {\vskip 3pt plus 0.1fill \beginlinemode
   \oneandahalfspace \raggedcenter \it}

\def\endtopmatter		
  {\endpage\global\pageno=1		
   \body}

\def\body			
  {\beginparmode}		

\def\beneathrel#1\under#2{\mathrel{\mathop{#2}\limits_{#1}}}


\def\ref#1{Ref.~#1}			
\def\Ref#1{Ref.~#1}			
\def\[#1]{[\cite{#1}]}
\def\cite#1{{#1}}
\def\(#1){(\call{#1})}
\def\call#1{{#1}}
\def\taghead#1{}
\def\frac#1#2{{#1 \over #2}}

\def\12{{1\over2}}

\def\etal{{\it et al.}}

\def\sla{\raise.15ex\hbox{$/$}\kern-.57em}
\def\leaderfill{\leaders\hbox to 1em{\hss.\hss}\hfill}
\def\twiddle{\lower.9ex\rlap{$\kern-.1em\scriptstyle\sim$}}
\def\bigtwiddle{\lower1.ex\rlap{$\sim$}}
\def\gtwid{\mathrel{\raise.3ex\hbox{$>$\kern-.75em\lower1ex\hbox{$\sim$}}}}
\def\ltwid{\mathrel{\raise.3ex\hbox{$<$\kern-.75em\lower1ex\hbox{$\sim$}}}}
\def\square{\kern1pt\vbox{\hrule height 1.2pt\hbox{\vrule width 1.2pt\hskip 3pt
   \vbox{\vskip 6pt}\hskip 3pt\vrule width 0.6pt}\hrule height 0.6pt}\kern1pt}
\def\tdot#1{\mathord{\mathop{#1}\limits^{\kern2pt\ldots}}}

\def\pmb#1{\setbox0=\hbox{#1}%
  \kern-.025em\copy0\kern-\wd0
  \kern  .05em\copy0\kern-\wd0
  \kern-.025em\raise.0433em\box0 }



\newcount\iiheadnumber
\newcount\iheadnumber
\newcount\headnumber
\newcount\tablenumber 
\newcount\figurenumber
\def\appendix#1&#2{ \endpage
	\vbox to 0.5truein{}	
	\vbox to 1.1truein{}	
	{\singlespace{\immediate\write16{#2}
	 \raggedcenter {\sc APPENDIX  #2} \par}
	{\immediate\write16{#1}
	\raggedcenter {\sc #1} \par}}
	\tocitem{#2\hskip .8cm #1}
        \nobreak\vskip 0.1truein\nobreak\global\headpage=\pageno
	\taghead{#2.}
        \xdef\chapterno{#2}
	\iheadnumber=0
	\headnumber=0
	\tablenumber=0
	\figurenumber=0
	\updownpages
}

\def\head#1{			
  \updownpages\tocitem{#1}
  \vbox to 0.9truein{}		
  {\immediate\write16{#1} 
   \raggedcenter {\sc #1} \par }
   \nobreak\vskip 0.1truein\nobreak\global\headpage=\pageno\beginparmode}

\def\chapter#1&#2{ \endpage
	\vbox to 1.1truein{}	
	{\singlespace{\immediate\write16{#2}
	 \raggedcenter {\sc CHAPTER  #2} \par}
	{\immediate\write16{#1}
	\raggedcenter {\sc #1} \par}}
	\tocitem{#2\hskip .8cm #1}
        \nobreak\vskip 0.1truein\nobreak\global\headpage=\pageno
        \xdef\chapterno{#2}
	\iheadnumber=0
	\headnumber=0
	\tablenumber=0
	\figurenumber=0
        \taghead{#2.}
	\downpages

}

\def\subheadnn#1{			
  \vskip 0.25truein		
  \line{\bf{ \fourteenrm\hss {#1}\hss\par}}
  \iiheadnumber=0
  \nobreak\vskip 0truein\nobreak}

\def\subhead#1{			
  \global\advance\iheadnumber by 1
  \vskip 0.25truein		
  \line{\bf{ \fourteenrm\hss \the\iheadnumber.\hskip .2cm {#1}\hss\par}}
  \iiheadnumber=0
  \nobreak\vskip 0truein\nobreak}

\def\subsubhead#1{		
  \global\advance\iiheadnumber by  1
  \vskip 0.25truein		
 {\noindent \chapterno.\the\iheadnumber.\the\iiheadnumber\hskip.2cm
	{\underbar{#1}}\par}     
  \tocitemitemitem{\chapterno.\the\iheadnumber.\the\iiheadnumber
    \hskip .2cm #1} 
  \nobreak\vskip 0truein\nobreak}

\def\appendix#1&#2{ \endpage
	\vbox to 0.5truein{}	
	\vbox to 1.1truein{}	
	{\singlespace{\immediate\write16{#2}
	 \raggedcenter {\sc APPENDIX  #2} \par}
	{\immediate\write16{#1}
	\raggedcenter {\sc #1} \par}}
	\tocitem{#2\hskip .8cm #1}
        \nobreak\vskip 0.1truein\nobreak\global\headpage=\pageno
	\taghead{#2.}
        \xdef\chapterno{#2}
	\iheadnumber=0
	\headnumber=0
	\tablenumber=0
	\figurenumber=0
	\updownpages
}


\def\newtable#1#2{\xdef#1{\chapterno.\the\tablenumber\relax}
  \centerline{{\bf Table} \the\tablenumber:\ {#2}}
  \vskip 6pt\nobreak
}

\def\TableTitle{%
     \global\advance\tablenumber by 1 
     \newtable}

\def\n,{\hskip -5pt ,}  
\def\n.{\hskip -5pt ,}  

\def\newfig#1#2#3{\xdef#1{\chapterno.\the\figurenumber\relax}
  \singlespace\noindent{\bf Figure \the\figurenumber:}\quad #2
  \vfil\vskip 10pt
  \noindent\singlespace#3
  \vfill
}

\def\newtit#1#2{\xdef#1{\chapterno.\the\figurenumber\relax}
  \singlespace\noindent{\bf Figure \ \the\figurenumber}\quad #2
  \vfil\vskip 10pt
  \vfill
  \vskip 20pt 
}

\def\figureCaption{%
     \global\advance\figurenumber by 1 
     \newfig}

\def\figureTitle{%
     \global\advance\figurenumber by 1 
     \newtit}

\def\figurecaptions		
  {\endpage
   \beginparmode
   \head{FIGURE CAPTIONS}
}


\def\refto#1{$^{#1}$}		

\def\references			
 {\line{ \bf{\fourteenrm\hss References \hss\par}}
   \beginparmode
   \parskip=\bigskipamount \singlespace 
	 \everypar{\hangindent=28pt\hangafter=1}}


\gdef\refis#1{\item{#1.\ }}			

\gdef\journal#1, #2,#3, #4 {			
    {\sl #1~}{\bf #2}, #3 (#4)}			

\def\refstyleprl{				
  \gdef\refto##1{ [##1]}			
 \gdef\refis##1{\item{[##1] }}			
  \gdef\journal##1, ##2, ##3 {	
    {##1~}{{\bf ##3},~}}}

\def\prd{\journal Phys. Rev. D, }

\def\pl{\journal Phys. Lett., }

\def\endreferences{\body}

\def\endpage			
  {\vfill\eject}

\def\endpaper			
  {\endmode\vfill\supereject}

\catcode`@=11
\newcount\tagnumber\tagnumber=0

\immediate\newwrite\eqnfile
\newif\if@qnfile\@qnfilefalse
\def\write@qn#1{}
\def\writenew@qn#1{}
\def\w@rnwrite#1{\write@qn{#1}\message{#1}}
\def\@rrwrite#1{\write@qn{#1}\errmessage{#1}}

\def\taghead#1{\gdef\t@ghead{#1}\global\tagnumber=0}
\def\t@ghead{}

\expandafter\def\csname @qnnum-3\endcsname
  {{\t@ghead\advance\tagnumber by -3\relax\number\tagnumber}}
\expandafter\def\csname @qnnum-2\endcsname
  {{\t@ghead\advance\tagnumber by -2\relax\number\tagnumber}}
\expandafter\def\csname @qnnum-1\endcsname
  {{\t@ghead\advance\tagnumber by -1\relax\number\tagnumber}}
\expandafter\def\csname @qnnum0\endcsname
  {\t@ghead\number\tagnumber}
\expandafter\def\csname @qnnum+1\endcsname
  {{\t@ghead\advance\tagnumber by 1\relax\number\tagnumber}}
\expandafter\def\csname @qnnum+2\endcsname
  {{\t@ghead\advance\tagnumber by 2\relax\number\tagnumber}}
\expandafter\def\csname @qnnum+3\endcsname
  {{\t@ghead\advance\tagnumber by 3\relax\number\tagnumber}}

\def\equationfile{%
  \@qnfiletrue\immediate\openout\eqnfile=\jobname.eqn%
  \def\write@qn##1{\if@qnfile\immediate\write\eqnfile{##1}\fi}
  \def\writenew@qn##1{\if@qnfile\immediate\write\eqnfile
    {\noexpand\tag{##1} = (\t@ghead\number\tagnumber)}\fi}
}

\def\callall#1{\xdef#1##1{#1{\noexpand\call{##1}}}}
\def\call#1{\each@rg\callr@nge{#1}}

\def\each@rg#1#2{{\let\thecsname=#1\expandafter\first@rg#2,\end,}}
\def\first@rg#1,{\thecsname{#1}\apply@rg}
\def\apply@rg#1,{\ifx\end#1\let\next=\relax%
\else,\thecsname{#1}\let\next=\apply@rg\fi\next}

\def\callr@nge#1{\calldor@nge#1-\end-}
\def\callr@ngeat#1\end-{#1}
\def\calldor@nge#1-#2-{\ifx\end#2\@qneatspace#1 %
  \else\calll@@p{#1}{#2}\callr@ngeat\fi}
\def\calll@@p#1#2{\ifnum#1>#2{\@rrwrite{Equation range #1-#2\space is bad.}
\errhelp{If you call a series of equations by the notation M-N, then M and
N must be integers, and N must be greater than or equal to M.}}\else%
 {\count0=#1\count1=#2\advance\count1 by1\relax\expandafter\@qncall\the\count0,%
  \loop\advance\count0 by1\relax%
    \ifnum\count0<\count1,\expandafter\@qncall\the\count0,%
  \repeat}\fi}

\def\@qneatspace#1#2 {\@qncall#1#2,}
\def\@qncall#1,{\ifunc@lled{#1}{\def\next{#1}\ifx\next\empty\else
  \w@rnwrite{Equation number \noexpand\(>>#1<<) has not been defined yet.}
  >>#1<<\fi}\else\csname @qnnum#1\endcsname\fi}

\let\eqnono=\eqno
\def\eqno(#1){\tag#1}
\def\tag#1$${\eqnono(\displayt@g#1 )$$}

\def\aligntag#1\endaligntag
  $${\gdef\tag##1\\{&(##1 )\cr}\eqalignno{#1\\}$$
  \gdef\tag##1$${\eqnono(\displayt@g##1 )$$}}

\def\eqalignno#1{\displ@y \tabskip\centering
  \halign to\displaywidth{\hfil$\displaystyle{##}$\tabskip\z@skip
    &$\displaystyle{{}##}$\hfil\tabskip\centering
    &\llap{$\displayt@gpar##$}\tabskip\z@skip\crcr
    #1\crcr}}

\def\displayt@gpar(#1){(\displayt@g#1 )}

\def\displayt@g#1 {\rm\ifunc@lled{#1}\global\advance\tagnumber by1
        {\def\next{#1}\ifx\next\empty\else\expandafter
        \xdef\csname @qnnum#1\endcsname{\t@ghead\number\tagnumber}\fi}%
  \writenew@qn{#1}\t@ghead\number\tagnumber\else
        {\edef\next{\t@ghead\number\tagnumber}%
        \expandafter\ifx\csname @qnnum#1\endcsname\next\else
        \w@rnwrite{Equation \noexpand\tag{#1} is a duplicate number.}\fi}%
  \csname @qnnum#1\endcsname\fi}

\def\ifunc@lled#1{\expandafter\ifx\csname @qnnum#1\endcsname\relax}

\let\@qnend=\end\gdef\end{\if@qnfile
\immediate\write16{Equation numbers written on []\jobname.EQN.}\fi\@qnend}

\catcode`@=12



\def\sigdbcd{\relax\ifmmode{ \sigma_{_{\rm DBCD}} }\else{ $\sigma_{_{\rm DBCD}}$\ }\fi}

\def\mGeV{\relax\ifmmode{{\rm GeV}}\else{${\rm GeV}$\ }\fi}
\def\mMeV{\relax\ifmmode{{\rm MeV}}\else{${\rm MeV}$\ }\fi}
\def\pGeV{\relax\ifmmode{{\rm GeV}}\else{${\rm GeV}$\ }\fi}
\def\pMeV{\relax\ifmmode{{\rm MeV}}\else{${\rm MeV}$\ }\fi}

\def\thrupork#1{\mathrel{\mathop{#1\!\!\!/}}}

\def\thru#1{\mathrel{\mathop{#1\!\!\!\!/}}}
\def\thrud#1{\mathrel{\mathop{#1\!\!\!\!/}}}

\def\gammu{\relax\ifmmode{\gamma^\mu}\else{$\gamma^\mu$\ }\fi}
\def\Amu{\relax\ifmmode{A_\mu}\else{$A_\mu$\ }\fi}
\def\Anu{\relax\ifmmode{A_\nu}\else{$A_\nu$\ }\fi}
\def\Dnu{\relax\ifmmode{D_\nu}\else{$D_\nu$\ }\fi}
\def\Dmu{\relax\ifmmode{D_\mu}\else{$D_\mu$\ }\fi}
\def\Fsq{\relax\ifmmode{F_{\mu\nu}F^{\mu\nu}}\else{$F_{\mu\nu}F^{\mu\nu}$\ }\fi}
\def\del{\relax\ifmmode{\partial}\else{$\partial$\ }\fi}

\def\delsl{\relax\ifmmode{\thrupork{\partial}}\else{$\thrupork{\partial}$\ }\fi}
\def\Asl{\relax\ifmmode{\thrupork{A}}\else{$\thrupork{A}$\ }\fi}
\def\Bsl{\relax\ifmmode{\thrud{B}}\else{$\thrud{B}$\ }\fi}
\def\Dsl{\relax\ifmmode{\thrud{D}}\else{$\thrud{D}$\ }\fi}
\def\Bsl{\relax\ifmmode{\thrud{B}}\else{$\thrud{B}$\ }\fi}

\def\delmu{\relax\ifmmode{\partial_\mu}\else{$\partial_\mu$\ }\fi}
\def\delnu{\relax\ifmmode{\partial_\nu}\else{$\partial_\nu$\ }\fi}
\def\psib{\relax\ifmmode{{\bar \psi}}\else{${\bar \psi}$\ }\fi}

\def\Bmu{\relax\ifmmode{B_\mu}\else{$B_\mu$\ }\fi}
\def\Bnu{\relax\ifmmode{B_\nu}\else{$B_\nu$\ }\fi}

\def\Wslmu{\relax\ifmmode{\thru{W}_\mu}\else{$\thru{W}_\mu$\ }\fi}
\def\Wmu{\relax\ifmmode{W_\mu}\else{$W_\mu$\ }\fi}
\def\Wslnu{\relax\ifmmode{\thrud{W}_\nu}\else{$\thrud{W}_\nu$\ }\fi}
\def\Wnu{\relax\ifmmode{W_\nu}\else{$W_\nu$\ }\fi}
\def\exp{\relax\ifmmode{{\rm e}}\else{${\rm e}$\ }\fi}

\def\SUiiL{\relax\ifmmode{SU(2)_L}\else{$SU(2)_L$\ }\fi}
\def\SUiiiC{\relax\ifmmode{SU(3)_C}\else{$SU(3)_C$\ }\fi}
\def\SUiif{\relax\ifmmode{SU(2)_f}\else{$SU(2)_f$\ }\fi}
\def\SUiiif{\relax\ifmmode{SU(3)_f}\else{$SU(3)_f$\ }\fi}
\def\SUiis{\relax\ifmmode{SU(2)_s}\else{$SU(2)_s$\ }\fi}
\def\SUivf{\relax\ifmmode{SU(4)_f}\else{$SU(4)_f$\ }\fi}
\def\SUiv{\relax\ifmmode{SU(4)}\else{$SU(4)$\ }\fi}
\def\UiY{\relax\ifmmode{U(1)_Y}\else{$U(1)_Y$\ }\fi}
\def\SUii{\relax\ifmmode{SU(2)}\else{$SU(2)$\ }\fi}
\def\SUiii{\relax\ifmmode{SU(3)}\else{$SU(3)$\ }\fi}
\def\Ui{\relax\ifmmode{U(1)}\else{$U(1)$\ }\fi}


\def\codefont{\tt}
\def\startline{\par\noindent}
{\obeylines\obeyspaces%
\global\def\beginCode{\displaybreak%
\begingroup%
\singlespace%
\parskip=0pt%
\obeylines\obeyspaces%
\let^^M=\startline%
\codefont}%
}

\def\displaybreak{\bigbreak}

\def\jhead#1#2{			
  \global\advance\headnumber by 1
  \updownpages\tocitem{#2}
  {\immediate\write16{#2} 
   \raggedcenter {\bf#1.\the\headnumber \hskip .5cm }{\bf {\sc #2}} \par }
   \nobreak\vskip 0.1truein\nobreak\global\headpage=\pageno\beginparmode}

\def\beginMidItems{\begingroup 
     \vskip 20pt
     \advance\parindent by 8em}

\def\beginLeftItems{\begingroup
     \vskip 20pt
     \advance\parindent by 2em}

%
%

\def\sigediff{\relax\ifmmode{\sigma_{\ediff}}\else{$\sigma_{\ediff}$\ }\fi}
\def\Ecm{\relax\ifmmode{E_{cm}}\else{$E_{cm}$\ }\fi}

\def\sig{\relax\ifmmode{\sigma}\else{$\sigma$\ }\fi}
\def\degr{\relax\ifmmode{^\circ}\else{$^\circ$}\fi}
\def\ediff{\relax\ifmmode{\Delta{\rm E}}\else{$\Delta{\rm E}$\ }\fi}
\def\bmass{\relax\ifmmode{{\rm M_{BC}}}\else{${\rm M_{BC}}$\ }\fi}
\def\spheri{\relax\ifmmode{\Theta_f}\else
	{$\Theta_f$\ }\fi}
\def\heli{\relax\ifmmode{\Theta_h}\else
	{$\Theta_h$\ }\fi}
\def\dmdiff{\relax\ifmmode{m_{D^{*}}-m_{D}}\else
	{${m_{D^{*}}-m_{D}}$\ }\fi}

\def\parone{\sl}

\def\tiny{\vrule width 0pt}
\def\star{{\bf *}}
\def\epem{\relax\ifmmode{e^+e^-}\else{$e^+e^-$\ }\fi}
\def\decays{\relax\ifmmode{\rightarrow}\else{$\rightarrow$\ }\fi\tiny}

\def\Vud{\relax\ifmmode{{\rm V}_{ud}}\else{{\rm V}$_{ud}$\ }\fi}
\def\Vcd{\relax\ifmmode{{\rm V}_{cd}}\else{{\rm V}$_{cd}$\ }\fi}
\def\Vtd{\relax\ifmmode{{\rm V}_{td}}\else{{\rm V}$_{td}$\ }\fi}
\def\Vus{\relax\ifmmode{{\rm V}_{us}}\else{{\rm V}$_{us}$\ }\fi}
\def\Vcs{\relax\ifmmode{{\rm V}_{cs}}\else{{\rm V}$_{cs}$\ }\fi}
\def\Vts{\relax\ifmmode{{\rm V}_{ts}}\else{{\rm V}$_{ts}$\ }\fi}
\def\Vub{\relax\ifmmode{{\rm V}_{ub}}\else{{\rm V}$_{ub}$\ }\fi}
\def\Vcb{\relax\ifmmode{{\rm V}_{cb}}\else{{\rm V}$_{cb}$\ }\fi}
\def\Vtb{\relax\ifmmode{{\rm V}_{tb}}\else{{\rm V}$_{tb}$\ }\fi}

\def\Dstrnostrp{\relax\ifmmode{{\parone D}^{(\star)+}}\else{${\parone D}^{(\star)+}$\ }\fi}
\def\Dstrnostrm{\relax\ifmmode{{\parone D}^{(\star)-}}\else{${\parone D}^{(\star)-}$\ }\fi}
\def\Dstrnostrz{\relax\ifmmode{{\parone D}^{(\star)0}}\else{${\parone D}^{(\star)0}$\ }\fi}

%
%


\def\gam{\relax\ifmmode{\gamma}\else{$\gamma$\ }\fi}


\def\W{\relax\ifmmode{{\parone W}}\else{{\parone W}}\fi}
\def\Wp{\relax\ifmmode{{\parone W}^+}\else{{\parone W}$^+$\ }\fi}
\def\Wm{\relax\ifmmode{{\parone W}^-}\else{{\parone W}$^-$\ }\fi}
\def\Wpm{\relax\ifmmode{{\parone W}^\pm}\else{{\parone W}$^\pm$\ }\fi}
\def\Wmp{\relax\ifmmode{{\parone W}^\mp}\else{{\parone W}$^\mp$\ }\fi}


\def\Z{\relax\ifmmode{{\parone Z}}\else{{\parone Z}}\fi}
\def\ZZ{\relax\ifmmode{{\parone Z}^0}\else{{\parone Z}$^0$\ }\fi}


\def\nub{\relax\ifmmode{\bar{\nu}}
	\else{$\bar{\nu}$\ }\fi}

\def\nue{\relax\ifmmode{\nu_e}\else{$\nu_e$\ }\fi}
\def\nueb{\relax\ifmmode{\bar{\nu}\tiny_e}
	\else{$\bar{\nu}\tiny_e$\ }\fi}


\def\e{\relax\ifmmode{e}\else{$e$\ }\fi}
\def\ep{\relax\ifmmode{e^+}\else{$e^+$\ }\fi}
\def\em{\relax\ifmmode{e^-}\else{$e^-$\ }\fi}
\def\epm{\relax\ifmmode{e^\pm}\else{$e^\pm$\ }\fi}
\def\emp{\relax\ifmmode{e^\mp}\else{$e^\mp$\ }\fi}


\def\numu{\relax\ifmmode{\nu_\mu}\else{$\nu_\mu$\ }\fi}
\def\numub{\relax\ifmmode{\bar{\nu}\tiny_\mu}
	\else{$\bar{\nu}\tiny_\mu$\ }\fi}


\def\nutau{\relax\ifmmode{\nu_\tau}\else{$\nu_\tau$\ }\fi}
\def\nutaub{\relax\ifmmode{\bar{\nu}\tiny_\tau}
	\else{$\bar{\nu}\tiny_\tau$\ }\fi}


\def\taup{\relax\ifmmode{\tau^+}\else{$\tau^+$\ }\fi}
\def\taum{\relax\ifmmode{\tau^-}\else{$\tau^-$\ }\fi}
\def\taupm{\relax\ifmmode{\tau^\pm}\else{$\tau^\pm$\ }\fi}
\def\taump{\relax\ifmmode{\tau^\mp}\else{$\tau^\mp$\ }\fi}

\def\pip{\relax\ifmmode{\pi^+}\else{$\pi^+$\ }\fi}
\def\piz{\relax\ifmmode{\pi^0}\else{$\pi^0$\ }\fi}
\def\pizs{\relax\ifmmode{\pi^0\rm{s}}\else{$\pi^0\rm{s}$\ }\fi}
\def\pim{\relax\ifmmode{\pi^-}\else{$\pi^-$\ }\fi}
\def\pipm{\relax\ifmmode{\pi^\pm}\else{$\pi^\pm$\ }\fi}
\def\pimp{\relax\ifmmode{\pi^\mp}\else{$\pi^\mp$\ }\fi}
\def\pipmz{\relax\ifmmode{\pi^{\pm,0}}\else{$\pi^{\pm,0}$\ }\fi}

\def\etap{\relax\ifmmode{\eta^{\prime}}\else{$\eta^{\prime}$\ }\fi}


\def\rhop{\relax\ifmmode{\rho^+}\else{$\rho^+$\ }\fi}
\def\rhoz{\relax\ifmmode{\rho^0}\else{$\rho^0$\ }\fi}
\def\rhom{\relax\ifmmode{\rho^-}\else{$\rho^-$\ }\fi}
\def\rhopm{\relax\ifmmode{\rho^\pm}\else{$\rho^\pm$\ }\fi}
\def\rhomp{\relax\ifmmode{\rho^\mp}\else{$\rho^\mp$\ }\fi}
\def\rhopmz{\relax\ifmmode{\rho^{\pm,0}}\else{$\rho^{\pm,0}$\ }\fi}

\def\omegz{\relax\ifmmode{\omega^0}\else{$\omega^0$\ }\fi}

\def\aone{\relax\ifmmode{a_{1}}\else{$a_{1}$\ }\fi}
\def\aonep{\relax\ifmmode{a^{+}_{1}}\else{$a^{+}_{1}$\ }\fi}
\def\aonez{\relax\ifmmode{a^{0}_{1}}\else{$a^{0}_{1}$\ }\fi}
\def\aonem{\relax\ifmmode{a^{-}_{1}}\else{$a^{-}_{1}$\ }\fi}
\def\aonepm{\relax\ifmmode{a^{\pm}_{1}}\else{$a^{\pm}_{_1}$\ }\fi}
\def\aonemp{\relax\ifmmode{a^{\mp}_{1}}\else{$a^{\mp}_{1}$\ }\fi}


\def\K{\relax\ifmmode{{\parone K}}\else{{\parone K}}\fi}
\def\Kb{\relax\ifmmode{\bar{{\parone K}}}
	\else{$\bar{{\parone K}}$\ }\fi}

\def\Kz{\relax\ifmmode{{\parone K}^0}\else{{\parone K}$^0$\ }\fi}
\def\Ksh{\relax\ifmmode{{\parone K}^0_S}\else{{\parone K}$^0_S$\ }\fi}
\def\Klo{\relax\ifmmode{{\parone K}^0_L}\else{{\parone K}$^0_L$\ }\fi}
\def\Kzb{\relax\ifmmode{\bar{{\parone K}}\tiny^0}
	\else{$\bar{{\parone K}}\tiny^0$\ }\fi}

\def\Kpm{\relax\ifmmode{{\parone K}^\pm/{\parone \pi}^\pm}\else{${\parone K}^\pm/{\parone \pi}^\pm$\ }\fi}

\def\Kp{\relax\ifmmode{{\parone K}^+}\else{{\parone K}$^+$\ }\fi}
\def\Km{\relax\ifmmode{{\parone K}^-}\else{{\parone K}$^-$\ }\fi}
\def\Kpm{\relax\ifmmode{{\parone K}^\pm}\else{{\parone K}$^\pm$\ }\fi}
\def\Kmp{\relax\ifmmode{{\parone K}^\mp}\else{{\parone K}$^\mp$\ }\fi}


\def\D{\relax\ifmmode{{\parone D}}\else{${\parone D}$\ }\fi}

\def\Db{\relax\ifmmode{\bar{{\parone D}}}\else{$\bar{{\parone D}}$\ }\fi}

\def\Dz{\relax\ifmmode{{\parone D}^0}\else{{\parone D}$^0$\ }\fi}
\def\Dzb{\relax\ifmmode{\bar{{\parone D}}\tiny^0}
	\else{$\bar{{\parone D}}\tiny^0$\ }\fi}

\def\Dp{\relax\ifmmode{{\parone D}^+}\else{{\parone D}$^+$\ }\fi}
\def\Dm{\relax\ifmmode{{\parone D}^-}\else{{\parone D}$^-$\ }\fi}
\def\Dpm{\relax\ifmmode{{\parone D}^\pm}\else{{\parone D}$^\pm$\ }\fi}
\def\Dmp{\relax\ifmmode{{\parone D}^\mp}\else{{\parone D}$^\mp$\ }\fi}


\def\prtn{\relax\ifmmode{{\parone p}}\else{{\parone p}}\fi}
\def\prtnb{\relax\ifmmode{\bar{{\parone p}}}
	\else{$\bar{{\parone p}}$\ }\fi}


\def\ntrn{\relax\ifmmode{{\parone n}}\else{{\parone n}}\fi}
\def\ntrnb{\relax\ifmmode{\bar{{\parone n}}}
	\else{$\bar{{\parone n}}$\ }\fi}


\def\lam{\relax\ifmmode{\Lambda}\else{$\Lambda$\ }\fi}
\def\lamb{\relax\ifmmode{\bar{\Lambda}}
	\else{$\bar{\Lambda}$\ }\fi}
\def\lamz{\relax\ifmmode{\Lambda^0}\else{$\Lambda^0$\ }\fi}
\def\lamzb{\relax\ifmmode{\bar{\Lambda}\tiny^0}
	\else{$\bar{\Lambda}\tiny^0$\ }\fi}


\def\US{\relax\ifmmode{\Upsilon(1S)}
			\else{$\Upsilon(1S)$\ }\fi}
\def\USS{\relax\ifmmode{\Upsilon(2S)}
	\else{$\Upsilon(2S)$\ }\fi}
\def\USSS{\relax\ifmmode{\Upsilon(3S)}
	\else{$\Upsilon(3S)$\ }\fi}
\def\USSSS{\relax\ifmmode{\Upsilon(4S)}
	\else{$\Upsilon(4S)$\ }\fi}
\def\USSSSS{\relax\ifmmode{\Upsilon(5S)}
	\else{$\Upsilon(5S)$\ }\fi}


\def\Kstr{\relax\ifmmode{{\parone K}^\star}\else{{\parone K}$^\star$\ }\fi}
\def\Kstrb{\relax\ifmmode{\b
ar{{\parone K}}\tiny^\star}
	\else{$\bar{{\parone K}}\tiny^\star$\ }\fi}

\def\Kstrz{\relax\ifmmode{{\parone K}^{\star0}}\else{{\parone K}$^{\star0}$\ }\fi}
\def\Kstrzb{\relax\ifmmode{\bar{{\parone K}}\tiny^{\star0}}
	\else{$\bar{{\parone K}}\tiny^{\star0}$\ }\fi}

\def\Kstrp{\relax\ifmmode{{\parone K}^{\star+}}\else{{\parone K}$^{\star+}$\ }\fi}
\def\Kstrm{\relax\ifmmode{{\parone K}^{\star-}}\else{{\parone K}$^{\star-}$\ }\fi}
\def\Kstrpm{\relax\ifmmode{{\parone K}^{\star\pm}}\else{{\parone K}$^{\star\pm}$\ }\fi}
\def\Kstrmp{\relax\ifmmode{{\parone K}^{\star\mp}}\else{{\parone K}$^{\star\mp}$\ }\fi}


\def\Dstr{\relax\ifmmode{{\parone D}^\star}\else{{\parone D}$^\star$\ }\fi}
\def\Dstrb{\relax\ifmmode{\bar{{\parone D}}\tiny^\star}
	\else{$\bar{{\parone D}}\tiny^\star$\ }\fi}

\def\Dstrz{\relax\ifmmode{{\parone D}^{\star 0}}\else{{\parone D}$^{\star 0}$\
}\fi}

\def\Dstrzb{\relax\ifmmode{\bar{{\parone D}}\tiny^{\star 0}}
	\else{$\bar{{\parone D}}\tiny^{\star 0}$\ }\fi}

\def\Dstrp{\relax\ifmmode{{\parone D}^{\star+}}\else{{\parone D}$^{\star+}$\ }\fi}
\def\Dstrm{\relax\ifmmode{{\parone D}^{\star-}}\else{{\parone D}$^{\star-}$\ }\fi}
\def\Dstrpm{\relax\ifmmode{{\parone D}^{\star\pm}}\else{{\parone D}$^{\star\pm}$\ }\fi}
\def\Dstrmp{\relax\ifmmode{{\parone D}^{\star\mp}}\else{{\parone D}$^{\star\mp}$\ }\fi}


\def\DDstr{\relax\ifmmode{{\parone D}^{\star\star}}\else{{\parone D}$^{\star\star}$\ }\fi}
\def\DDstrb{\relax\ifmmode{\bar{{\parone D}}\tiny^{\star\star}}
	\else{$\bar{{\parone D}}\tiny^{\star\star}$\ }\fi}

\def\DDstrz{\relax\ifmmode{{\parone D}^{\star\star0}}\else{{\parone D}$^{\star\star0}$\ }\fi}
\def\DDstrzb{\relax\ifmmode{\bar{{\parone D}}\tiny^{\star\star0}}
	\else{$\bar{{\parone D}}\tiny^{\star\star0}$\ }\fi}

\def\DDstrp{\relax\ifmmode{{\parone D}^{\star\star+}}\else{{\parone D}$^{\star\star+}$\ }\fi}
\def\DDstrm{\relax\ifmmode{{\parone D}^{\star\star-}}\else{{\parone D}$^{\star\star-}$\ }\fi}
\def\DDstrpm{\relax\ifmmode{{\parone D}^{\star\star\pm}}
	\else{{\parone D}$^{\star\star\pm}$\ }\fi}
\def\DDstrmp{\relax\ifmmode{{\parone D}^{\star\star\mp}}
	\else{{\parone D}$^{\star\star\mp}$\ }\fi}

\def\B{\relax\ifmmode{\parone{B}}\else{$\parone{B}$\ }\fi}
\def\Bb{\relax\ifmmode{\bar{\parone{B}}}\else{$\bar{{\parone B}}$\ }\fi}

\def\Bz{\relax\ifmmode{{\parone B}^0}\else{${\parone B}^0$\ }\fi}
\def\Bzb{\relax\ifmmode{\bar{\parone B}^0}\else{$\bar{\parone B}^0$\ }\fi}

\def\Bp{\relax\ifmmode{\parone{B}^+}\else{$\parone{B}^+$\ }\fi}
\def\Bm{\relax\ifmmode{\parone{B}^-}\else{$\parone {B}^-$\ }\fi}

\def\bbar{\relax\ifmmode{B\bar{B}}\else{$B\bar{B}$\ }\fi}
\def\qqbar{\relax\ifmmode{q\bar{q}}\else{$q\bar{q}$\ }\fi}

%
%

\def\dzpiz{\relax\ifmmode{\Dstrz \to \Dz \piz}\else
	{$\Dstrz \to \Dz \piz$\ }\fi}
\def\dzpip{\relax\ifmmode{D^{*+} \to D^0\pi^+}\else 
	{$D^{*+} \to D^0\pi^+$\ }\fi}

\def\dzgam{\relax\ifmmode{\Dstrz \to \Dz \gamma}\else
	{$\Dstrz \to \Dz \gamma$\ }\fi}

\def\Kmpip{\relax\ifmmode{\Dz \to \Km \pip}\else
	{$\Dz \to \Km \pip$\ }\fi}
\def\Kmpippiz{\relax\ifmmode{\Dz \to \Km \pip \piz}\else
	{$\Dz \to \Km \pip \piz$\ }\fi}
\def\Kmpippimpip{\relax\ifmmode{\Dz \to \Km \pip \pim \pip}\else
	{$\Dz \to \Km \pip \pim \pip$\ }\fi}
\def\Kmpippip{\relax\ifmmode{\Dp \to \Km \pip \pip}\else
	{$\Dp \to \Km \pip \pip$\ }\fi}

\def\Kpi{\relax\ifmmode{\Dz \to \Km \pip}\else
	{$\Dz \to \Km \pip$\ }\fi}
\def\Kpipiz{\relax\ifmmode{\Dz \to \Km \pip \piz}\else
	{$\Dz \to \Km \pip \piz$\ }\fi}
\def\Kpipipi{\relax\ifmmode{\Dz \to \Km \pip \pim \pip}\else
	{$\Dz \to \Km \pip \pim \pip$\ }\fi}
\def\Kpipi{\relax\ifmmode{\Dp \to \Km \pip \pip}\else
	{$\Dp \to \Km \pip \pip$\ }\fi}

%
%
\def\decrhop{\relax\ifmmode{\rhop \to \pip \piz}\else
	{$\rhop \to \pip \piz$\ }\fi}
\def\decomega{\relax\ifmmode{\omega \to \pip \pim \piz}\else
	{$\omega \to \pip \pim \piz$\ }\fi}
\def\deceta{\relax\ifmmode{\eta \to \gamm \gamm}\else
	{$\eta \to \gamm \gamm$\ }\fi}
\def\decetap{\relax\ifmmode{\etap \to \eta \gamm \gamm}\else
	{$\etap \to \eta \gamm \gamm$\ }\fi}

%
%

\def\Dpi{\relax\ifmmode{\Bzb \to \Dp \pim}\else
	{$\Bzb \to \Dp \pim$\ }\fi}
\def\Drho{\relax\ifmmode{\Bzb \to \Dp \rhom}\else
	{$\Bzb \to \Dp \rhom$\ }\fi}
\def\Daone{\relax\ifmmode{\Bzb \to \Dp \aonem}\else
    	{$\Bzb \to \Dp \aonem$\ }\fi}

\def\Dstrpi{\relax\ifmmode{\Bzb \to \Dstrp \pim}\else
	{$\Bzb \to \Dstrp \pim$\ }\fi}
\def\Dstrrho{\relax\ifmmode{\Bzb \to \Dstrp\rhom}\else
	{$\Bzb \to \Dstrp\rhom$\ }\fi}
\def\Dstraone{\relax\ifmmode{\Bzb \to \Dstrp \aonem}\else
	{$\Bzb \to \Dstrp \aonem$\ }\fi}

\def\Dzpiz{\relax\ifmmode{\Bzb \to \Dz \piz}\else
	{$\Bzb \to \Dz \piz$\ }\fi}
\def\Dzrhoz{\relax\ifmmode{\Bzb \to \Dz \rhoz}\else
	{$\Bzb \to \Dz \rhoz$\ }\fi}
\def\Dzeta{\relax\ifmmode{\Bzb \to \Dz \eta}\else
	{$\Bzb \to \Dz \eta$\ }\fi}
\def\Dzetap{\relax\ifmmode{\Bzb \to \Dz \etap}\else
	{$\Bzb \to \Dz \etap$\ }\fi}
\def\Dzomega{\relax\ifmmode{\Bzb \to \Dz \omega}\else
	{$\Bzb \to \Dz \omega$\ }\fi}
\def\Dzaonez{\relax\ifmmode{\Bzb \to \Dz \aonez}\else
	{$\Bzb \to \Dz \aonez$\ }\fi}

\def\Dstrzpiz{\relax\ifmmode{\Bzb \to \Dstrz \piz}\else
	{$\Bzb \to \Dstrz \piz$\ }\fi}
\def\Dstrzrhoz{\relax\ifmmode{\Bzb \to \Dstrz \rhoz}\else
	{$\Bzb \to \Dstrz \rhoz$\ }\fi}
\def\Dstrzeta{\relax\ifmmode{\Bzb \to \Dstrz \eta}\else
	{$\Bzb \to \Dstrz \eta$\ }\fi}
\def\Dstrzetap{\relax\ifmmode{\Bzb \to \Dstrz \etap}\else
	{$\Bzb \to \Dstrz \etap$\ }\fi}
\def\Dstrzomega{\relax\ifmmode{\Bzb \to \Dstrz \omega}\else
	{$\Bzb \to \Dstrz \omega$\ }\fi}
\def\Dstrzaonez{\relax\ifmmode{\Bzb \to \Dstrz \aonez}\else
	{$\Bzb \to \Dstrz \aonez$\ }\fi}

\def\Dzpi{\relax\ifmmode{\Bm \to \Dz \pim}\else
	{$\Bm \to \Dz \pim$\ }\fi}
\def\Dzrho{\relax\ifmmode{\Bm \to \Dz \rhom}\else
	{$\Bm \to \Dz \rhom$\ }\fi}
\def\Dzaone{\relax\ifmmode{\Bm \to \Dz \aonem}\else
	{$\Bm \to \Dz \aonem$\ }\fi}

\def\Dstrzpi{\relax\ifmmode{\Bm \to \Dstrz \pim}\else
	{$\Bm \to \Dstrz \pim$\ }\fi}
\def\Dstrzrho{\relax\ifmmode{\Bm \to \Dstrz \rhom}\else
	{$\Bm \to \Dstrz \rhom$\ }\fi}
\def\Dstrzaone{\relax\ifmmode{\Bm \to \Dstrz \aonem}\else
	{$\Bm \to \Dstrz \aonem$\ }\fi}

%
%
\def\Dstrzgpi{\relax\ifmmode{\Bm \to \Dstrz \pim}\else
	{$\Bm \to \Dstrz \pim$\ }\fi}
\def\Dstrzrho{\relax\ifmmode{\Bm \to \Dstrz \rhom}\else
	{$\Bm \to \Dstrz \rhom$\ }\fi}


%% file: reforder.tex
\catcode`@=11
\newcount\r@fcount \r@fcount=0
\newcount\r@fcurr
\immediate\newwrite\reffile
\newif\ifr@ffile\r@ffilefalse
\def\w@rnwrite#1{\ifr@ffile\immediate\write\reffile{#1}\fi\message{#1}}

\def\writer@f#1>>{}
\def\referencefile{
  \r@ffiletrue\immediate\openout\reffile=\jobname.ref%
  \def\writer@f##1>>{\ifr@ffile\immediate\write\reffile%
    {\noexpand\refis{##1} = \csname r@fnum##1\endcsname = %
     \expandafter\expandafter\expandafter\strip@t\expandafter%
     \meaning\csname r@ftext\csname r@fnum##1\endcsname\endcsname}\fi}%
  \def\strip@t##1>>{}}

\def\citeall#1{\xdef#1##1{#1{\noexpand\cite{##1}}}}
\def\cite#1{\each@rg\citer@nge{#1}}	

\def\each@rg#1#2{{\let\thecsname=#1\expandafter\first@rg#2,\end,}}
\def\first@rg#1,{\thecsname{#1}\apply@rg}	
\def\apply@rg#1,{\ifx\end#1\let\next=\relax
\else,\thecsname{#1}\let\next=\apply@rg\fi\next}

\def\citer@nge#1{\citedor@nge#1-\end-}	
\def\citer@ngeat#1\end-{#1}
\def\citedor@nge#1-#2-{\ifx\end#2\r@featspace#1 
  \else\citel@@p{#1}{#2}\citer@ngeat\fi}	
\def\citel@@p#1#2{\ifnum#1>#2{\errmessage{Reference range #1-#2\space is bad.}
    \errhelp{If you cite a series of references by the notation M-N, then M and
    N must be integers, and N must be greater than or equal to M.}}\else%
 {\count0=#1\count1=#2\advance\count1 by1\relax\expandafter\r@fcite\the\count0,%
  \loop\advance\count0 by1\relax
    \ifnum\count0<\count1,\expandafter\r@fcite\the\count0,%
  \repeat}\fi}

\def\r@featspace#1#2 {\r@fcite#1#2,}	
\def\r@fcite#1,{\ifuncit@d{#1}
    \expandafter\gdef\csname r@ftext\number\r@fcount\endcsname%
    {\message{Reference #1 to be supplied.}\writer@f#1>>#1 to be supplied.\par
     }\fi%
  \csname r@fnum#1\endcsname}

\def\ifuncit@d#1{\expandafter\ifx\csname r@fnum#1\endcsname\relax%
\global\advance\r@fcount by1%
\expandafter\xdef\csname r@fnum#1\endcsname{\number\r@fcount}}

\let\r@fis=\refis			
\def\refis#1#2#3\par{\ifuncit@d{#1}
    \w@rnwrite{Reference #1=\number\r@fcount\space is not cited up to now.}\fi%
  \expandafter\gdef\csname r@ftext\csname r@fnum#1\endcsname\endcsname%
  {\writer@f#1>>#2#3\par}}

\def\r@ferr{\endreferences\errmessage{I was expecting to see
\noexpand\endreferences before now;  I have inserted it here.}}
\let\r@ferences=\references
\def\references{\r@ferences\def\endmode{\r@ferr\par\endgroup}}

\let\endr@ferences=\endreferences
\def\endreferences{\r@fcurr=0
  {\loop\ifnum\r@fcurr<\r@fcount
    \advance\r@fcurr by 1\relax\expandafter\r@fis\expandafter{\number\r@fcurr}%
    \csname r@ftext\number\r@fcurr\endcsname%
  \repeat}\gdef\r@ferr{}\endr@ferences}


\let\r@fend=\endpaper\gdef\endpaper{\ifr@ffile
\immediate\write16{Cross References written on []\jobname.REF.}\fi\r@fend}

\catcode`@=12

\citeall\refto		
\citeall\ref		%
\citeall\Ref		%

%% file: tables.tex
%
\newbox\hdbox%
\newcount\hdrows%
\newcount\multispancount%
\newcount\ncase%
\newcount\ncols
\newcount\nrows%
\newcount\nspan%
\newcount\ntemp%
\newdimen\hdsize%
\newdimen\newhdsize%
\newdimen\parasize%
\newdimen\spreadwidth%
\newdimen\thicksize%
\newdimen\thinsize%
\newdimen\tablewidth%
\newif\ifcentertables%
\newif\ifendsize%
\newif\iffirstrow%
\newif\iftableinfo%
\newtoks\dbt%
\newtoks\hdtks%
\newtoks\savetks%
\newtoks\tableLETtokens%
\newtoks\tabletokens%
\newtoks\widthspec%
%
%
\immediate\write15{%
CP SMSG GJMSINK TEXTABLE --> TABLE MACROS V. 851121 JOB = \jobname%
}%
%
%
\tableinfotrue%
\catcode`\@=11
%
%
\def\tstrut{\vrule height3.1ex depth1.2ex width0pt}%
\def\and{\char`\&}
\def\tablerule{\noalign{\hrule height\thinsize depth0pt}}%
\thicksize=1.5pt
\thinsize=0.6pt
\def\thickrule{\noalign{\hrule height\thicksize depth0pt}}%
\def\ctr#1{\hfil\ #1\hfil}%
%
%
%
%
\tablewidth=-\maxdimen%
\spreadwidth=-\maxdimen%
\def\tabskipglue{0pt plus 1fil minus 1fil}%
%
%
\centertablestrue%
%
%
%
%
\parasize=4in%
\gdef\ARGS{########}
\gdef\headerARGS{####}
\def\@mpersand{&}
{\catcode`\|=13
\gdef\letbarzero{\let|0}
\gdef\letbartab{\def|{&&}}%
\gdef\letvbbar{\let\vb|}%
}
{\catcode`\&=4
\def\ampskip{&\omit\hfil&}
\catcode`\&=13
\let&0
\xdef\letampskip{\def&{\ampskip}}%
\gdef\letnovbamp{\let\novb&\let\tab&}
}
\def\begintable{
   \begingroup%
   \catcode`\|=13\letbartab\letvbbar%
   \catcode`\&=13\letampskip\letnovbamp%
   \def\multispan##1{
      \omit \mscount##1%
      \multiply\mscount\tw@\advance\mscount\m@ne%
      \loop\ifnum\mscount>\@ne \sp@n\repeat%
   }
   \def\|{%
      &\omit\widevline&%
   }%
   \ruledtable
}
\long\def\ruledtable#1\endtable{%
%
%
%
   \offinterlineskip
   \tabskip 0pt
   \def\widevline{\vrule width\thicksize}
   \def\endrow{\@mpersand\omit\hfil\crnorm\@mpersand}%
   \def\crthick{\@mpersand\crnorm\thickrule\@mpersand}%
   \def\crthickneg##1{\@mpersand\crnorm\thickrule
          \noalign{{\skip0=##1\vskip-\skip0}}\@mpersand}%
   \def\crnorule{\@mpersand\crnorm\@mpersand}%
   \def\crnoruleneg##1{\@mpersand\crnorm
          \noalign{{\skip0=##1\vskip-\skip0}}\@mpersand}%
   \let\nr=\crnorule
   \def\endtable{\@mpersand\crnorm\thickrule}%
   \let\crnorm=\cr
%
%
   \edef\cr{\@mpersand\crnorm\tablerule\@mpersand}%
   \def\crneg##1{\@mpersand\crnorm\tablerule
          \noalign{{\skip0=##1\vskip-\skip0}}\@mpersand}%
   \let\ctneg=\crthickneg
   \let\nrneg=\crnoruleneg
   \the\tableLETtokens
%
%
   \tabletokens={&#1}
%
%
   \countROWS\tabletokens\into\nrows%
   \countCOLS\tabletokens\into\ncols%
%
%
   \advance\ncols by -1%
   \divide\ncols by 2%
   \advance\nrows by 1%
%
%
   \iftableinfo %
      \immediate\write16{[Nrows=\the\nrows, Ncols=\the\ncols]}%
   \fi%
%
%
   \ifcentertables
      \ifhmode \par\fi
      \line{
      \hss
   \else %
      \hbox{%
   \fi
      \vbox{%
         \makePREAMBLE{\the\ncols}
         \edef\next{\preamble}
         \let\preamble=\next
         \makeTABLE{\preamble}{\tabletokens}
      }
      \ifcentertables \hss}\else }\fi
   \endgroup
   \tablewidth=-\maxdimen
   \spreadwidth=-\maxdimen
}
\def\makeTABLE#1#2{
   {
   \let\ifmath0
   \let\header0
   \let\multispan0
%
%
   \ncase=0%
   \ifdim\tablewidth>-\maxdimen \ncase=1\fi%
   \ifdim\spreadwidth>-\maxdimen \ncase=2\fi%
   \relax
%
   \ifcase\ncase %
      \widthspec={}%
   \or %
      \widthspec=\expandafter{\expandafter t\expandafter o%
                 \the\tablewidth}%
   \else %
      \widthspec=\expandafter{\expandafter s\expandafter p\expandafter r%
                 \expandafter e\expandafter a\expandafter d%
                 \the\spreadwidth}%
   \fi %
   \xdef\next{
      \halign\the\widthspec{%
      #1
      \noalign{\hrule height\thicksize depth0pt}
      \the#2\endtable
%
      }
   }
   }
   \next
}
\def\makePREAMBLE#1{
   \ncols=#1
   \begingroup
   \let\ARGS=0
   \edef\xtp{\widevline\ARGS\tabskip\tabskipglue%
   &\ctr{\ARGS}\tstrut}
   \advance\ncols by -1
   \loop
      \ifnum\ncols>0 %
      \advance\ncols by -1%
      \edef\xtp{\xtp&\vrule width\thinsize\ARGS&\ctr{\ARGS}}%
   \repeat
   \xdef\preamble{\xtp&\widevline\ARGS\tabskip0pt%
   \crnorm}
   \endgroup
}
\def\countROWS#1\into#2{
   \let\countREGISTER=#2%
   \countREGISTER=0%
   \expandafter\ROWcount\the#1\endcount%
}%
\def\ROWcount{%
   \afterassignment\subROWcount\let\next= %
}%
\def\subROWcount{%
   \ifx\next\endcount %
      \let\next=\relax%
   \else%
      \ncase=0%
      \ifx\next\cr %
         \global\advance\countREGISTER by 1%
         \ncase=0%
      \fi%
      \ifx\next\endrow %
         \global\advance\countREGISTER by 1%
         \ncase=0%
      \fi%
      \ifx\next\crthick %
         \global\advance\countREGISTER by 1%
         \ncase=0%
      \fi%
      \ifx\next\crnorule %
         \global\advance\countREGISTER by 1%
         \ncase=0%
      \fi%
      \ifx\next\crthickneg %
         \global\advance\countREGISTER by 1%
         \ncase=0%
      \fi%
      \ifx\next\crnoruleneg %
         \global\advance\countREGISTER by 1%
         \ncase=0%
      \fi%
      \ifx\next\crneg %
         \global\advance\countREGISTER by 1%
         \ncase=0%
      \fi%
      \ifx\next\header %
         \ncase=1%
      \fi%
      \relax%
      \ifcase\ncase %
         \let\next\ROWcount%
      \or %
         \let\next\argROWskip%
      \else %
      \fi%
   \fi%
   \next%
}
\def\counthdROWS#1\into#2{%
\dvr{10}%
   \let\countREGISTER=#2%
   \countREGISTER=0%
\dvr{11}%
\dvr{13}%
   \expandafter\hdROWcount\the#1\endcount%
\dvr{12}%
}%
\def\hdROWcount{%
   \afterassignment\subhdROWcount\let\next= %
}%
\def\subhdROWcount{%
   \ifx\next\endcount %
      \let\next=\relax%
   \else%
      \ncase=0%
      \ifx\next\cr %
         \global\advance\countREGISTER by 1%
         \ncase=0%
      \fi%
      \ifx\next\endrow %
         \global\advance\countREGISTER by 1%
         \ncase=0%
      \fi%
      \ifx\next\crthick %
         \global\advance\countREGISTER by 1%
         \ncase=0%
      \fi%
      \ifx\next\crnorule %
         \global\advance\countREGISTER by 1%
         \ncase=0%
      \fi%
      \ifx\next\header %
         \ncase=1%
      \fi%
\relax%
      \ifcase\ncase %
         \let\next\hdROWcount%
      \or%
         \let\next\arghdROWskip%
      \else %
      \fi%
   \fi%
   \next%
}%
{\catcode`\|=13\letbartab
\gdef\countCOLS#1\into#2{%
   \let\countREGISTER=#2%
   \global\countREGISTER=0%
   \global\multispancount=0%
   \global\firstrowtrue
   \expandafter\COLcount\the#1\endcount%
   \global\advance\countREGISTER by 3%
   \global\advance\countREGISTER by -\multispancount
}%
\gdef\COLcount{%
   \afterassignment\subCOLcount\let\next= %
}%
{\catcode`\&=13%
\gdef\subCOLcount{%
   \ifx\next\endcount %
      \let\next=\relax%
   \else%
      \ncase=0%
      \iffirstrow
         \ifx\next& %
            \global\advance\countREGISTER by 2%
            \ncase=0%
         \fi%
         \ifx\next\span %
            \global\advance\countREGISTER by 1%
            \ncase=0%
         \fi%
         \ifx\next| %
            \global\advance\countREGISTER by 2%
            \ncase=0%
         \fi
         \ifx\next\|
            \global\advance\countREGISTER by 2%
            \ncase=0%
         \fi
         \ifx\next\multispan
            \ncase=1%
            \global\advance\multispancount by 1%
         \fi
         \ifx\next\header
            \ncase=2%
         \fi
         \ifx\next\cr       \global\firstrowfalse \fi
         \ifx\next\endrow   \global\firstrowfalse \fi
         \ifx\next\crthick  \global\firstrowfalse \fi
         \ifx\next\crnorule \global\firstrowfalse \fi
         \ifx\next\crnoruleneg \global\firstrowfalse \fi
         \ifx\next\crthickneg  \global\firstrowfalse \fi
         \ifx\next\crneg       \global\firstrowfalse \fi
      \fi
\relax
      \ifcase\ncase %
         \let\next\COLcount%
      \or %
         \let\next\spancount%
      \or %
         \let\next\argCOLskip%
      \else %
      \fi %
   \fi%
   \next%
}%
\gdef\argROWskip#1{%
   \let\next\ROWcount \next%
}
\gdef\arghdROWskip#1{%
   \let\next\ROWcount \next%
}
\gdef\argCOLskip#1{%
   \let\next\COLcount \next%
}
}
}
\def\spancount#1{
   \nspan=#1\multiply\nspan by 2\advance\nspan by -1%
   \global\advance \countREGISTER by \nspan
   \let\next\COLcount \next}%
\def\dvr#1{\relax}%
\def\header#1{%
\dvr{1}{\let\cr=\@mpersand%
\hdtks={#1}%
\counthdROWS\hdtks\into\hdrows%
\advance\hdrows by 1%
\ifnum\hdrows=0 \hdrows=1 \fi%
\dvr{5}\makehdPREAMBLE{\the\hdrows}%
\dvr{6}\getHDdimen{#1}%
{\parindent=0pt\hsize=\hdsize{\let\ifmath0%
\xdef\next{\valign{\headerpreamble #1\crnorm}}}\dvr{7}\next\dvr{8}%
}%
}\dvr{2}}
\def\makehdPREAMBLE#1{
\dvr{3}%
\hdrows=#1
{
\let\headerARGS=0%
\let\cr=\crnorm%
\edef\xtp{\vfil\hfil\hbox{\headerARGS}\hfil\vfil}%
\advance\hdrows by -1
\loop
\ifnum\hdrows>0%
\advance\hdrows by -1%
\edef\xtp{\xtp&\vfil\hfil\hbox{\headerARGS}\hfil\vfil}%
\repeat%
\xdef\headerpreamble{\xtp\crcr}%
}
\dvr{4}}
\def\getHDdimen#1{%
\hdsize=0pt%
\getsize#1\cr\end\cr%
}
\def\getsize#1\cr{%
\endsizefalse\savetks={#1}%
\expandafter\lookend\the\savetks\cr%
\relax \ifendsize \let\next\relax \else%
\setbox\hdbox=\hbox{#1}\newhdsize=1.0\wd\hdbox%
\ifdim\newhdsize>\hdsize \hdsize=\newhdsize \fi%
\let\next\getsize \fi%
\next%
}%
\def\lookend{\afterassignment\sublookend\let\looknext= }%
\def\sublookend{\relax%
\ifx\looknext\cr %
\let\looknext\relax \else %
   \relax
   \ifx\looknext\end \global\endsizetrue \fi%
   \let\looknext=\lookend%
    \fi \looknext%
}%
%
%
\def\tablelet#1{%
   \tableLETtokens=\expandafter{\the\tableLETtokens #1}%
}%
\catcode`\@=12

%% file: tbl_normall.tex
\TableTitle\tblnormall{Branching fraction measurements for class I and
III decays} 

{\begintable

$\B^0$ Mode    & $\cal B (\%)$ ||$\B^-$ Mode    & $\cal B (\%)$ \cr

$\Dp\pim$      \hfil&$0.308\pm 0.026\pm 0.028\pm 0.031\ $||$\Dz\pim$      \hfil&$0.534\pm 0.025\pm 0.033\pm 0.047\ $\crnorule 
		                                                                                                        
$\Dp\rhom$     \hfil&$0.861\pm 0.078\pm 0.109\pm 0.086\ $||$\Dstrz\pim$   \hfil&$0.497\pm 0.044\pm 0.048\pm 0.057\ $\crnorule  
						                                                          
$\Dstrp\pim$   \hfil&$0.304\pm 0.024\pm 0.025\pm 0.027\ $||$\Dz\rhom$     \hfil&$1.022\pm 0.067\pm 0.109\pm 0.086\ $\crnorule
	                                     	                                                          
$\Dstrp\rhom$  \hfil&$0.844\pm 0.071\pm 0.096\pm 0.076\ $||$\Dstrz\rhom$  \hfil&$1.444\pm 0.134\pm 0.188\pm 0.161\ $\crnorule  
$\Dstrp\aonem$ \hfil&$1.205\pm 0.140\pm 0.138\pm 0.098\ $||$\Dstrz\aonem$ \hfil&$1.898\pm 0.268\pm 0.236\pm 0.221\ $ 
\endtable}

%% file: tbl_factestbr.tex
\TableTitle\tblfactestbr{Tests of factorization} 

{\begintable
 $q^2$ |\hskip 10pt $R_{exp} \ ({\rm GeV}^2)$\hskip 10pt|\hskip 5pt $R_{th} \ ({\rm GeV}^2)$ \hskip 5pt\cr
$m_{\pi}^2$ | $1.3 \pm 0.1 \pm 0.2$| $1.2\pm 0.2$ \crnorule
$m_\rho^2$ | $3.4 \pm 0.3 \pm 0.5$| $3.3\pm 0.5$ \crnorule
$m_{a_1}^2$| $3.8 \pm 0.4 \pm 0.5$| $3.0\pm 0.5$ 

\endtable}

%% file: tbl_color_all.tex
\TableTitle\tblcolorall{Branching fraction limits for class II
(color-suppressed) decays @ 90\% C.L.}
{\begintable

Decay Mode  & \ \ ${\rm N_{obs}}$\ \ \ &  ${\cal B}$ (\%) ||Decay Mode  & \ \ ${\rm N_{obs}}$\ \ \ &  ${\cal B}$ (\%) \cr

\Dzpiz      \hfil & $ < 33.3 $\hfil &$ <0.033\ $||\Dstrzpiz   \hfil & $ < 14.6 $\hfil &$ <0.055 $    \crnorule
\Dzeta      \hfil & $ <  9.4 $\hfil &$ <0.033\ $||\Dstrzeta   \hfil & $ <  3.6 $\hfil &$ <0.050 $    \crnorule
\Dzetap     \hfil & $ <  2.3 $\hfil &$ <0.029\ $||\Dstrzetap  \hfil & $ <  2.3 $\hfil &$ <0.13  $    \crnorule
\Dzrhoz     \hfil & $ < 33.7 $\hfil &$ <0.060\ $||\Dstrzrhoz  \hfil & $ < 19.1 $\hfil &$ <0.12  $    \crnorule
\Dzomega    \hfil & $ < 13.0 $\hfil &$ <0.057\ $||\Dstrzomega \hfil & $ < 11.8 $\hfil &$ <0.12  $    

\endtable}